\documentclass[aps,prb,longbibliography,amsmath,amssymb,superscriptaddress,twocolumn]{revtex4-2}
\usepackage[utf8]{inputenc}
\usepackage{graphicx}
\usepackage{mathrsfs}
\usepackage{mathtools}
\usepackage{bm}
\usepackage{textcomp,xcolor}
\usepackage[colorlinks=true,urlcolor=blue,citecolor=blue,linkcolor=blue,bookmarks=false, pdfstartview={FitH}]{hyperref}
\usepackage[capitalize]{cleveref} 
\usepackage{amsmath}
\usepackage[normalem]{ulem} 

\newcommand{\beq}{\begin{equation}}
  \newcommand{\eeq}{\end{equation}}
\newcommand{\bea}{\begin{eqnarray}}
  \newcommand{\eea}{\end{eqnarray}}
\long\def\bal#1\eal{\begin{align}#1\end{align}}
\newcommand{\bse}{\begin{subequations}}
  \newcommand{\ese}{\end{subequations}}
\newcommand{\nn}{\nonumber}
\newcommand{\bwt}{\begin{widetext}}
  \newcommand{\ewt}{\end{widetext}}

\newcommand{\ve}{\varepsilon}

\usepackage{xparse}
\DeclareDocumentCommand\differential{ o g d() }{%
  \IfNoValueTF{#2}{
    \IfNoValueTF{#3}
    {\mathrm{d}\IfNoValueTF{#1}{}{^{#1}}}
    {\mathinner{\mathrm{d}\IfNoValueTF{#1}{}{^{#1}}\argopen(#3\argclose)}}
  }
  {\mathinner{\mathrm{d}\IfNoValueTF{#1}{}{^{#1}}#2} \IfNoValueTF{#3}{}{(#3)}}
}
\DeclareDocumentCommand\dd{}{\differential}
\newcommand{\ii}{\ensuremath{{\mkern1mu\mathrm{i}\mkern1mu}}}

\DeclareMathOperator{\IM}{Im}
\usepackage[version=3]{mhchem}
\newcommand{\group}[1]{\ce{#1}}
\newcommand{\Bog}{\group{B_{1g}}}
\newcommand{\Btg}{\group{B_{2g}}}
\newcommand{\Aog}{\group{A_{1g}}}

\newcommand{\customcref}[2]{\namecref{#1}~\hyperref[#1]{\ref{#1}#2}}
\newcommand{\customcrefs}[2]{\namecrefs{#1}~\hyperref[#1]{\ref{#1}#2}}

\begin{document}

\title{A modified-residue prescription to calculate dynamical correlation functions}
\author{Igor Benek-Lins}
\affiliation{Department of Physics, Concordia University, Montréal, QC H4B 1R6, Canada}
\author{Jonathan Discenza}
\affiliation{Department of Physics, Concordia University, Montréal, QC H4B 1R6, Canada}
\author{Saurabh Maiti}
\affiliation{Department of Physics, Concordia University, Montréal, QC H4B 1R6, Canada}
\affiliation{Centre for Research in Multiscale Modelling, Concordia University, Montréal, QC H4B 1R6, Canada}
\date{\today}
\begin{abstract}
  One of the challenges in using numerical methods to address many-body problems is the multi-dimensional integration over poles. More often that not, one needs such integrations to be evaluated as a function of an external variable. An example would be calculating dynamical correlations functions that are used to model response functions, where the external variable is the frequency. The standard numerical techniques rely on building an adaptive mesh, using special points in the Brillouin zone or using advanced smearing techniques. Most of these techniques, however, suffer when the grid is coarse. Here we propose that, if one knows the nature of the singularity in the integrand, one can define a residue and use it to faithfully estimate the integral and reproduce all the resulting singular features even with a coarse grid. We demonstrate the effectiveness of the method for different scenarios of calculating correlation functions with different resulting singular features, for calculating collective modes and densities of states. We also present a quantitative analysis of the error and show that this method can be widely applicable.
\end{abstract}

\maketitle
\tableofcontents

\section{Introduction}\label{Sec:Introduction}
A response from a $D$-dimensional quantum system ($D\in\{1,2,3\}$), quite generally, involves integration over their Brillouin zones (BZs). In metallic systems, the integration region is often restricted to the Fermi surface, which still forms a $D-1$ dimensional manifold (for $D\ge 2$). Most of these integrations involve integration of singularities resulting from a pole or a line/surface of poles, which makes the calculation computationally very expensive, especially if one also needs to account for many-body effects~\cite{bruus_flensberg:2004:ManybodyQuantumTheoryCondensed,altland_simons:2023:CondensedMatterFieldTheory,mahan:2011:CondensedMatterNutshell,kent_kotliar:2018:PredictiveTheoryCorrelatedMaterials}. This problem has been long recognized and has led to many prescriptions that help with the cost of computation without compromising on the result's accuracy. These include exploiting special points in the BZ~\cite{chadi_cohen:1973:SpecialPointsBrillouinZone,monkhorst_pack:1976:SpecialPointsBrillouinzoneIntegrations}, using the tetrahedron method over BZ points~\cite{blochl_etal:1994:ImprovedTetrahedronMethodBrillouinzone,kawamura_etal:2014:ImprovedTetrahedronMethodBrillouinzone,assmann_etal:2016:WopticOpticalConductivityWannier,Guterding2018,cances_etal:2020:NumericalQuadratureBrillouinZone,Ghim2022}, using adaptive mesh in the BZ~\cite{henk:2001:IntegrationTwodimensionalBrillouinZones,kaye_etal:2023:AutomaticHighorderAdaptiveAlgorithms,vanmunoz_etal:2024:HighorderAdaptiveOpticalConductivity}, and using advanced smearing/broadening methods~\cite{methfessel_paxton:1989:HighprecisionSamplingBrillouinzoneIntegration,yates_etal:2007:SpectralFermiSurfaceProperties,dossantos_marzari:2023:FermiEnergyDeterminationAdvanced}, to name a few.

When studying response functions, there are two main types of singularities of interest. One is that in the integrand, and the other is that in the integral. The singular -- or nonanalytic -- features that result from the integration also carries useful information about the properties of the system~\cite{phillips:1956:CriticalPointsLatticeVibration,phillips:1957:ErratumCriticalPointsLattice,phillips:1966:FundamentalOpticalSpectraSolids,benek-lins_etal:2024:UniversalNonanalyticFeaturesResponse}. In many cases, the resulting singular features are the result of many-body effects and are not easily captured using the prescriptions listed above.
The major issue is that all of these approaches help improve \emph{ab initio} methods.

When calculating a two-body spectral function of a new quantum ground state, such as a superconductor, the universal low-energy properties are not directly dependent on the original lattice and BZ parameters. In such problems, the computation of the ground state itself is so expensive that one has to coarsely discretize the Fermi surface and the computation is done using a fixed `grid' of values. Techniques based on functional renormalization group (fRG)~\cite{salmhofer_honerkamp:2001:FermionicRenormalizationGroupFlows,kopietz_etal:2010:IntroductionFunctionalRenormalizationGroup,metzner_etal:2012:FunctionalRenormalizationGroupApproach} and random-phase approximation (RPA)~\cite{graser_etal:2009:NeardegeneracySeveralPairingChannels,reuther_wolfle:2010:J_1J_2FrustratedTwodimensional} are good examples that illustrate this.

This discretization makes the subsequent computation of a dynamical response function (that is a function of an external frequency $\omega$) not amenable to the automated adaptive algorithms as there is no longer the freedom to add more intermediate grid points without re-doing the already expensive many-body ground-state calculation. In order to be able to model responses from such materials, it would be beneficial to have a method that can accurately handle the singularities of the integrand and reproduce singular features of the resulting integral, all with the original coarse mesh used to obtain the ground state.

This is what is addressed in this work. We prescribe a method to calculate these dynamical response functions from an integrand that is only `available' at points of a coarse grid. We emphasize that in condensed-matter systems, the new ground state can be non-commensurate with the lattice~\cite{Schlottmann2016,Baggari2018} and, hence, it is not possible to choose the special points of the BZ \emph{a priori}, as required by some of the above methods. To be definite, we are interested in computing integrals of the type
\begin{equation}\label{eq:form0}
  \chi(\omega) = \int\frac{\dd[D]{k}}{(2\pi)^D}I(\vec{k},\omega),
\end{equation}
where $D$ is the spatial dimension, the $k$-integration is over the BZ, and the integrand $I(\vec{k},\omega)$ contains a line or a surface of poles at different $\vec{k}$ values. We will assume that $I(\vec{k},\omega)$ is only available at a coarse set of $\vec{k}$-points.

Our solution to the problem of accuracy stems from recognizing the nature of the singularity in the integrand. It has been known that it is possible to extract the asymptotic form of the singularity of the integrand~\cite{phillips:1956:CriticalPointsLatticeVibration,phillips:1957:ErratumCriticalPointsLattice,benek-lins_etal:2024:UniversalNonanalyticFeaturesResponse} in multidimensional integrals. This singularity is usually not a simple pole. However, analogous to an integration around a simple pole, we use the knowledge of the asymptotic singularity to introduce an integration scheme based on the `residue' of this modified pole. We call this the \emph{modified-residue prescription}. The modified-residue serves as the weight needed to sum over the coarse set of points to reproduce an accurate result. We demonstrate the validity of the method by calculating several response functions of anisotropic superconductors that are relevant to Raman scattering and their density of states. In particular, we show that it provides an order-of-magnitude improvement in the error when compared with a Riemann sum over the $\vec{k}$ points (which is often the only resort). We argue that if one knows the nature of the singularity, this prescription can be effectively incorporated with other general integration methods such as the method of quadratures.

In Sec.~\ref{Sec:setup}, we derive and state the modified-residue prescription and demonstrate its validity with a quick example. In Sec.~\ref{Sec:Examples}, we demonstrate the use of our prescription on spectral functions that are used in the Raman response of superconductors, in a related many-body collective mode problem and in the calculation of the density of states. In Sec.~\ref{Sec:ErrorAnalysis}, we quantify the error of the prescription and explain an interesting characteristic of our prescription: it outperforms the standard routines up to a minimum threshold size of the grid. We explain this behaviour and show that the threshold is at an impractically small grid size, rendering this prescription useful in almost all practical scenarios. Finally, in Sec.~\ref{Sec:Conclusion}, we summarize our findings and discusses the scope for future use of this method. In \cref{sec:appendix}, we discuss the specific form of the residue that is used in the text.

\bigskip

\section{The modified-residue prescription}\label{Sec:setup}
Let us start with a function $\chi(\omega)$ that is obtained after a 1D integration:
\begin{equation}\label{eq:Intgrand}
  \chi(\omega) = \int_{a}^{b}\dd{x} I(\omega, x),
\end{equation}
where the integrand $I$ has an integrable singularity at $x=x^*\in[a,b)$. Since such a singularity is fast varying, it can be well approximated by a local asymptotic form $s(\omega,y)$ such that $I(\omega,x)=R(\omega,x)s(\omega,x-x^*)$ in the neighbourhood of the singularity at $x=x^*$. This local asymptotic form can always be deduced, even if the integrand itself is obtained from a separate integration (see, e.g., \cite{benek-lins_etal:2024:UniversalNonanalyticFeaturesResponse}, which discusses such a method). The value $R(\omega,x^*)$ is nothing but $\lim_{x\rightarrow x^*} I(\omega,x)/s(\omega,x-x^*)$ and can be called as the `residue' of the singularity $s(\omega,y)$. This is a generalization of the conventional residue from a simple pole where $s(y)=1/y$ and we refer to it as a `modified residue'.

We can then identify $R(\omega,x)\equiv I(\omega,x)/s(\omega,x-x^{*})$ as the (modified) residue function that takes on the value of the residue at $x=x^*$. Of course, note that \(R(\omega,x)\), by construction, is a slow varying function relative to $s(\omega,x-x^{*})$ near $x=x^*$. With these new definitions, we can then evaluate \cref{eq:Intgrand} as
\begin{widetext}
\begin{align}\label{eq:Rform3}
  \chi(\omega) &=\int_a^{x^*-\delta/2}\dd{x} I(\omega,x) +\int^{x^*+\delta/2}_{x^*-\delta/2}\dd{x} \underbrace{I(\omega,x)}_{R(\omega,x)s(\omega,x-x^*)}+ \int^b_{x^*+\delta/2}\dd{x} I(\omega,x)\nn\\
  &\approx\int_a^{x^*-\delta/2}\dd{x} I(\omega,x) +R(\omega,x^*)\underbrace{\int^{x^*+\delta/2}_{x^*-\delta/2}\dd{x} s(\omega,x-x^*)}_{S(\omega,x^*)}+ \int^b_{x^*+\delta/2}\dd{x} I(\omega,x),
\end{align}
\end{widetext}
where $\delta$ is a suitable value that controls the size of the sub-interval over which the singular part is to be integrated over. In the second line, we have used the fact that the residue $R(\omega,x)$ is slowly varying and pulled it out of the integration, replacing it with its value at \(x^{*}\).

In the non-singular sub-intervals, the integral itself, $I(\omega,x)$, is slow-varying and can be approximated by usual integration routines which require discretizing the interval. In doing so, it would be beneficial to choose the size of the sub-intervals to be $\delta$ so that the singular sub-interval is naturally accounted for in the partitioning of the integration interval $[a,b)$.

For definiteness, we will use the Riemann sum as the conventional prescription. Therefore, if we break the integration into $N$ equal sub-intervals of size \(\delta\), we can write
\begin{align}\label{eq:Rform}
  &\chi(\omega)\nn\\
  &= \sum_{m=1}^{N}\int_{x_{m-1}}^{x_{m}}\dd{x} I(\omega, x),\nn\\
  &=\sum_{\substack{m=1 \\ m\neq m^*}}^{N}I(\omega, \bar x_m)\underbrace{(x_m-x_{m-1})}_{\delta}+R(\omega,\bar x_{m^*}) S(\omega,\bar x_{m^*}),
\end{align}
where \((x_0, x_N)=(a, b)\), $\bar x_m$ is a point in the sub-interval $[x_{m-1},x_m)$ chosen according to some given prescription (e.g., the mid-point of the interval) and $m^*$ is the index corresponding to the singular sub-interval.

Next, note that for non-singular sub-intervals
\begin{align}\label{eq:note1}
I(\omega,\bar x_m)\delta &= \frac{I(\omega,\bar x_m)}{s(\omega,\bar x_m-x^*)}s(\omega,\bar x_m-x^*)\delta\nn\\
&= R(\omega,\bar x_m)s(\omega,\bar x_m-x^*)\delta\nn\\
&\approx R(\omega,\bar x_m)\underbrace{\int_{x_{m-1}}^{x_m}\dd{x} s(\omega,x-x^*)}_{\mathclap{S_m(\omega)~\equiv~S(\omega,x_{m})~-~S(\omega,x_{m-1})}},
\end{align}
where the functional $S$ is the integral of $s(x)$. The last line uses the fact that in a non-singular sub-interval, even $s(\omega,y)$ is slowly varying, allowing us the replace $s\delta$ with an integral of \(s\) over the interval of size $\delta$.

We thus see that we can express the integral over the entire interval as a weighted sum of the modified-residue over all the sub-intervals. That is,
\begin{align}\label{eq:Rform2}
  \chi(\omega)\approx\chi^{\rm mRes}(\omega) &\equiv \sum_{m=1}^{N}R(\omega, \bar x_m)S_m(\omega) \nn\\
  &=\sum_{m=1}^{N}\frac{I(\omega, \bar x_m)}{s(\omega, \bar x_m)}S_m(\omega).
\end{align}
This is what we call the modified-residue (mRes) prescription.

In practice, $I(\omega,\bar x_m)$ would be the sample points that are numerically made available to us that we need to integrate over.
More often than not, unfortunately, the $m$-mesh is not fine enough to successfully deploy the Riemann sum (RSum) prescription to integrate over the singular nature of $I(\omega,x)$ [which would mean approximating $\chi(\omega)$ with $\chi^{\text{RSum}}(\omega)=\sum_m I(\omega,\bar x_m)\delta$]. However, if we knew the asymptotic singular form $s(\omega,x-x^*)$, we could compute $S(\omega,x)$ analytically. Then, using its definition in \cref{eq:note1}, we can evaluate $S_m(\omega)$ and deploy \cref{eq:Rform2} to estimate $\chi(\omega)$. As is evident from this construction, for a fixed $N$, \(\chi^{\text{mRes}}\) should perform the same as \(\chi^{\text{RSum}}\) in the non-singular sub-intervals and improve on it in the singular sub-interval, rendering an overall improvement in the former's performance.

\subsection{Demonstration of the prescription}

\begin{figure*}
  \centering
  \includegraphics[width=1\linewidth]{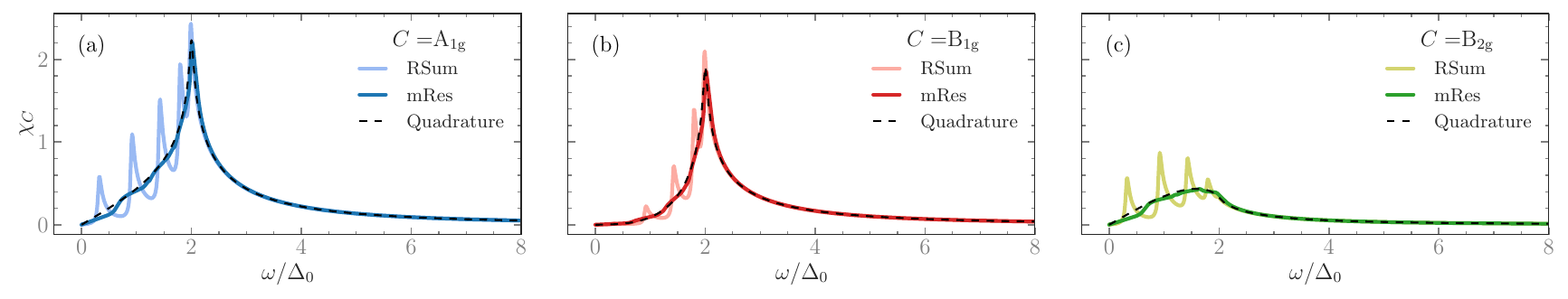}
  \caption{\textbf{Superiority of the modified-residue prescription.} A comparison of the Riemann sum (RSum) and the modified-residue prescription (mRes) to evaluate $\chi_C(\omega)$, with $\Delta_\theta=\Delta_0\cos(2\theta)$ for $C\in\{\Aog, \Bog, \Btg\}$. Here $N=40$. The exact calculations, obtained using the method of quadratures (which uses a high resolution and adaptive mesh), are shown as dashed lines.}\label{fig:ils_example}
\end{figure*}
As a quick proof of concept, let us demonstrate the above claim for the integral
\begin{widetext}
\begin{equation}\label{eq:demo}
  \chi_C(\omega)=\IM\Bigg[\lim_{\eta\rightarrow0}\int_0^{2\pi}\frac{\dd{\theta}}{2\pi}~\underbrace{\gamma^2_C(\theta)\int_{-\infty}^{\infty} \dd{\ve} \frac{\Delta_\theta^2}{\sqrt{\ve^2+\Delta_\theta^2}}\frac1{\ve^2+\Delta_\theta^2-(\omega+\ii\eta)^2/4}}_{I(\omega,\theta)}\Bigg],
\end{equation}
\end{widetext}
where $\Delta_\theta=\Delta_0\cos(2\theta)$, with $\Delta_0>0$, $C\in\{\Aog, \Bog, \Btg\}$ and $\gamma_{\Aog}(\theta)=1$, $\gamma_{\Bog}(\theta)=\cos(2\theta)$, $\gamma_{\Btg}(\theta)=\sin(2\theta)$. The meaning and relevance of these terms and symbols are not relevant at this stage (but will be elaborated on in the next section). It suffices to know that they just give us access to different scenarios where the applicability of the mRes prescription can be demonstrated.

Equation~(\ref{eq:demo}) represents a common scenario in many-body physics problems: we need the integration over a variable (here $\theta$), but the integrand $I(\omega,\theta)$ itself is a result of a different, and potentially involved, numerical calculation. Thus, the availability of $I(\omega,\theta)$ for different values of $\theta$ is usually limited. To use our prescription, we would need to know the resulting singularity after performing the 1D integration.

For this demonstrative example, the 1D integration is exactly doable leading to
\begin{align}\label{eq:I1}
  I(\omega,\theta)=2\gamma_C^2(\theta)\frac{\arcsin[\omega/(2|\Delta_\theta|)]}{\omega/(2|\Delta_\theta|)}\frac{|\Delta_\theta|}{\sqrt{\Delta^2_\theta-(\omega+\ii\eta)^2/4}}.
\end{align}
The singular part $\sqrt{\Delta_\theta^2-(\omega+\ii\eta)^2/4}$ in the denominator allows us to identify the local asymptotic form $s(\omega,\theta)$ as $1/\sqrt{(\omega+\ii\eta)/(2|\Delta_\theta|) - 1}$. The precise representation of the singularity ($1/\sqrt{x^2-1}$ or $1/\sqrt{x-1}$ forms near $x\sim1$) is not important as long as $S_m(\omega)$ uses the appropriate $s(\omega,x)$ -- see Appendix \ref{sec:appendix}. Although such closed-form results are not always achievable, one does not need to explicitly evaluate the integral to deduce \(s(\omega, \theta)\)~\cite{benek-lins_etal:2024:UniversalNonanalyticFeaturesResponse}.

Knowing $I(\omega,\theta)$ and $s(\omega,\theta)$, we can now follow the prescription in \cref{eq:Rform2}. In \cref{fig:ils_example} we plot the `exact' result with the dashed line. It was obtained using an adaptive quadrature numerical routine (SciPy's \emph{nquad}~\cite{scipy1_0contributors_etal:2020:SciPyFundamentalAlgorithmsScientific}) on Eq.~(\ref{eq:I1}).
We then present the result for the modified-residue prescription with $N=40$ with dark solid lines and that with the Riemann sum with light solid lines. The panels (a)--(c) correspond to different choices of $C$ which affects the slope of the low-$\omega$ behaviour and also the singularity (or lack thereof) at $\omega=2\Delta_0$. It is evident that the residue sum outperforms the Riemann sum and correctly captures the slopes and singular features of the result. This improvement will be quantified in Sec. \ref{Sec:ErrorAnalysis}.

\subsection{Incorporation with other prescriptions}

The Riemann sum discussed above is also referred to as the rectangle rule for integration. Consider also the following prescriptions~\cite{riley_etal:2006:MathematicalMethodsPhysicsEngineering} which are known to perform better than the rectangle rule:
\begin{itemize}
  \item Trapezoidal rule, which implements $$\int_a^b \dd{x} I(x)\approx\frac{\delta}2\Big[I(a)+2\sum_{m=1}^{N-1}I(x_m)+I(b)\Big].$$
  \item Simpson's rule, which implements $$\int_a^b \dd{x} I(x)\approx\frac{\delta}3\Big[I(a)+4\sum_{\mathclap{\text{odd } m}}I(x_m)+2\sum_{\mathclap{\text{even } m}}I(x_m)+I(b)\Big].$$
  \item Gaussian quadratures, which rescales the integration limits and the function to evaluate $$\int_{-1}^{1} \dd{x} I(x)\approx\sum_mw_mI(x_m),$$ for some pre-computed sample points $\{x_m\}$ and $\{w_m\}$ (according to some orthogonal set of polynomials) for a given $N$.
\end{itemize}

In all of these prescriptions, we have essentially a weighted sum over the discrete points $I(x_m)$. The weights, however, are not tied to any form of a singularity and, therefore, finer meshes would be demanded to correctly integrate singular integrands using these prescriptions.
Nonetheless, our prescription could be incorporated into the above schemes to obtain better results without requiring a finer mesh by suitably promoting $\delta$ to $S_m/s(x_m)$, provided we have knowledge of the nature of the local asymptotic singularity $s(x)$. The logic remains that in the non-singular sub-intervals, the proposed form already implements the original prescription, while providing the appropriate value for the singular sub-integral.

\paragraph*{Note:} It is worth noting that $S_m$ can also be evaluated numerically with a high density of grid points. This is not the same as evaluating $\int \dd{\theta} I(\omega,\theta)$ numerically with the same high density of grid points because we do not have access to $I(\theta)$ at those finer grid points. But knowing the asymptotic form, albeit only in the singular sub-interval, provides an analytical structure that can be integrated over a finer grid. This step is not really more computationally expensive that using the analytical form for $S_m$.
At this point, we may also note that when one needs to compute multi-dimensional integrals, one often resorts to Monte Carlo-based techniques. While these are known to perform best when the integrand is bounded, there are special algorithms designed to deal with singular integrands (see, for instance, Ref.~\cite{Atannasov2008}). We leave the incorporation of this idea into Monte Carlo techniques to a future endeavour.

\section{Examples}\label{Sec:Examples}
\begin{figure*}
  \centering
  \includegraphics[width=0.9\linewidth]{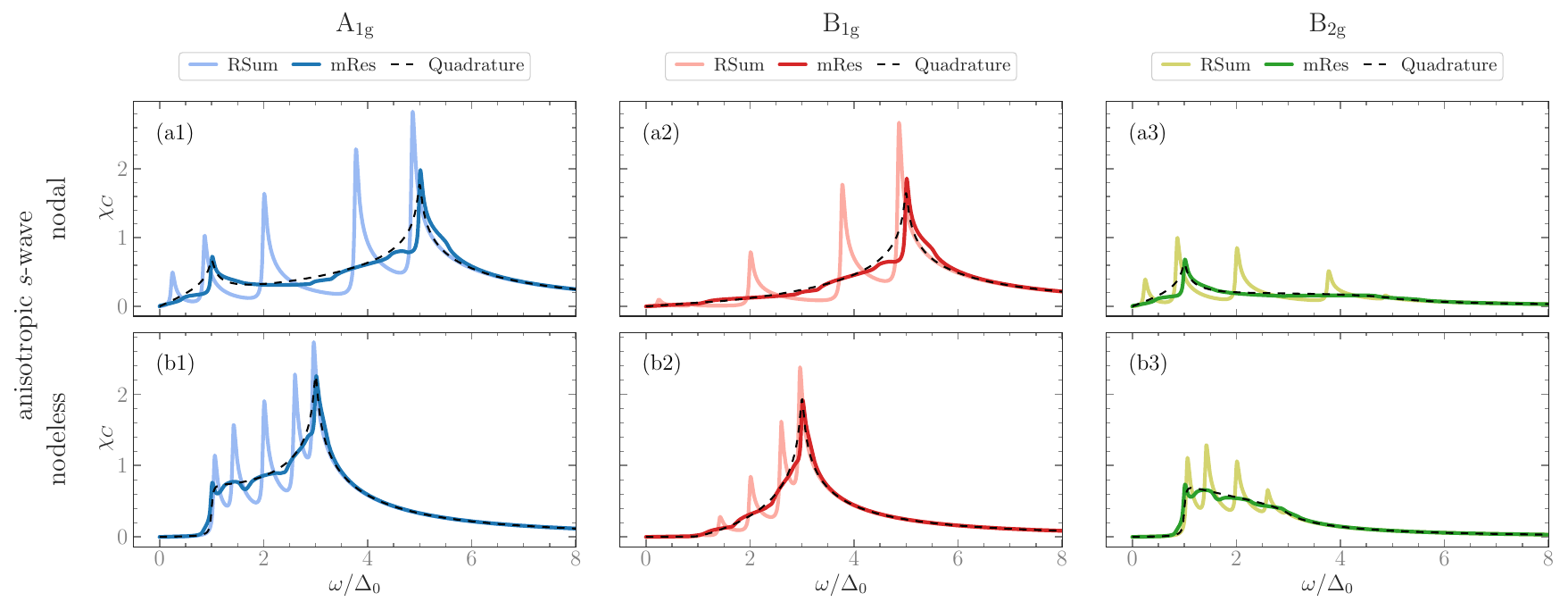}
  \caption{\textbf{The residue prescription and different singularities.} A comparison of RSum and mRes for $\chi_C(\omega)$, with $\Delta_\theta=\Delta_0+\Delta_1\cos(4\theta)$ for $C\in\{\Aog, \Bog, \Btg\}$ and the same number of subdivisions, $N=40$.
    The exact calculation, obtained from quadrature, is shown as dashed lines.
    The top panel is for the nodal anisotropic $s$-wave case, $\Delta_1>\Delta_0$, whilst the bottom panel is for the nodeless case, $\Delta_1<\Delta_0$.
    The prescription is capable of effectively reproducing the many different types of singularities.}\label{fig:ils_results}
\end{figure*}
In this section, we demonstrate some practical examples where we show the applicability and the improved performance of this prescription. As before, in all these cases, the exact calculation is performed using an adaptive quadrature routine.
We then compare how well the results computed using the RSum and the mRes methods converge to it for a fixed number of sub-intervals.
The comparisons will be between the Riemann sum and the residue prescription. While it should be sufficient to stick to the results of Fig.~\ref{fig:ils_example} and move on to error analysis, we think that it is beneficial to show how convincingly our method correctly captures the slopes, jumps and singularities of the integrated results in diverse scenarios.

\subsection{Spectral functions}
\begin{figure}[!b]
  \centering
  \includegraphics[width=1\linewidth]{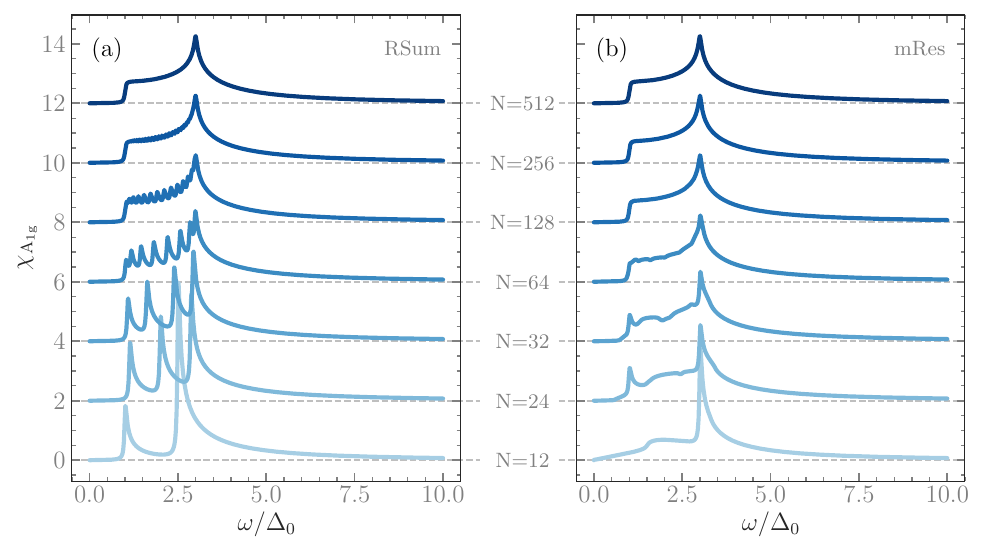}
  \caption{\textbf{Faster convergence of mRes.} Side-by-side comparison of the (a) RSum and (b) mRes prescription for different numbers of subdivisions $N$. We see that the latter converges already for a very low (even an order of magnitude lower) value of $N$. Here, we show the nodeless anisotropic \(s\)-wave case and the responses are shifted such that the dashed horizontal lines mark the zeros.}\label{fig:ils_comparison}
\end{figure}
\begin{figure*}[!ht]
  \centering
  \includegraphics[width=1.0\linewidth]{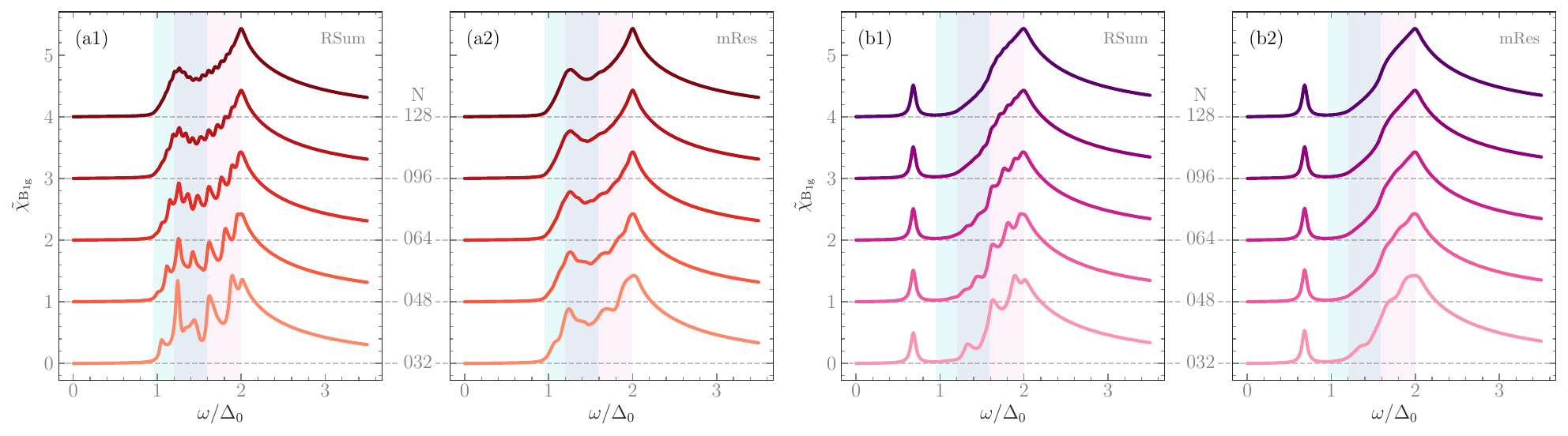}
  \caption{\textbf{Collective-mode scenario.}
    Side-by-side comparison of the Raman response of a two-band nodeless anisotropic \(s\)-wave superconductor when accounting for many-body effects in scenarios where a Bardasis--Schrieffer phase collective mode is present. These manifest as sharp features in response functions.
    Panels (a) correspond to a situation where we do not have any sharp collective modes as is evidenced by the large-$N$ plots, while the panels (b) correspond to a situation where we do have a sharp collective mode. As is evident, in the low-$N$ scenario it is difficult to disentangle spurious peaks from peaks due to collective modes in the RSum prescriptions -- (a1) and (b1), but is more readily done in mRes prescription -- (a2) and (b2). The shaded regions represents the spread of anisotropy of the order parameters.}\label{fig:b1a1_mode}
\end{figure*}
Let us return to the evaluation of \cref{eq:demo}. We chose this function as it is sufficiently complex to display various spectral features in terms of (i) different power laws in the limit $\omega\rightarrow0$, (ii) the presence or absence of singular peaks at various $\omega$'s and (iii) even the presence of discontinuous step jumps. In this subsection, we will demonstrate that our prescription covers all these different singular cases and faithfully reproduces the expected curve while significantly outperforming the rectangle rule for the same mesh size.

Let us also take this opportunity to explain the physical nature of the response function $\chi_C(\omega)$, Eq.~(\ref{eq:demo}). This function captures the Raman scattering cross-section from a 2D superconductor with order parameter $\Delta_\theta$. The choices for $C$ represent the orthogonal channels in which the contributions from the fermionic scattering can be broken down into. Different channels can be accessed by appropriately selecting the polarization of the incident and scattered light in the Raman experiment. The response $\chi_C$, by itself, is insufficient to reproduce the real physical response as this expression does not account for the many-body effects. Nevertheless, it forms a building block of the full many-body-corrected response and, hence, it becomes important to compute $\chi_C$ accurately. We refer the reader to the extensive literature on this subject for more information~\cite{klein_dierker:1984:TheoryRamanScatteringSuperconductors,monien_zawadowski:1990:TheoryRamanScatteringFinalstate,devereaux_hackl:2007:InelasticLightScatteringCorrelated, cea_benfatto:2016:SignatureLeggettModeGroup,maiti_etal:2017:ConservationLawsVertexCorrections}.

The choice of $\Delta_\theta=\Delta_0\cos(2\theta)$, that was used in Fig.~\ref{fig:ils_example}, actually corresponds to a so-called $d$-wave superconductor (best example being the cuprate superconductors~\cite{shen_davis:2008:CuprateHighT_mathrmcSuperconductors,tsuei_kirtley:2000:PairingSymmetryCuprateSuperconductors}). However, we have other materials such as the Fe-based pnictides and chalcogenides~\cite{chubukov_hirschfeld:2015:IronbasedSuperconductorsSevenYears,fernandes_etal:2022:IronPnictidesChalcogenidesNew}, where the structure of the order parameter may take the form $\Delta_\theta=\Delta_0+\Delta_1\cos4\theta$. This order-parameter structure allows for two scenarios: (i) $\Delta_1>\Delta_0$, which is referred to as the \emph{nodal} case where the order parameter reaches zero at certain angles; and (ii) $\Delta_1<\Delta_0$, which is referred to as the \emph{nodeless} case, as the order parameter is finite for all angles, but is anisotropic. These are referred to as the anisotropic $s$-wave order parameters. The form of $I(\omega,\theta)$ and $s(\omega,\theta)$ associated with them remain the same as before, but with the use of the appropriate $\Delta_\theta$. In \cref{fig:ils_results}, we show the comparison of the RSum and the mRes methods for different choices of $C\in\{\Aog,\Bog,\Btg\}$, which changes the number and the location of the singular features. The panels \hyperref[fig:ils_results]{(a)} correspond to the nodal order parameter and the panels \hyperref[fig:ils_results]{(b)} to the nodeless one.

In Fig.~\ref{fig:ils_comparison}, we present the scenario from \customcref{fig:ils_results}{(b1)} but comparing the RSum and the mRes prescription for different values of $N$. The faster convergence of the residue prescription is evident.

\subsection{Collective modes}
Owing to the potential high cost of increasing the number of sub-intervals, it is beneficial to keep this number low. However, as we saw in Fig.~\ref{fig:ils_comparison}, we run into the risk of introducing spurious peaks in the response that will misguide the inference drawn from the calculation. We can demonstrate this issue by performing a calculation for a simple model that calculates the full Raman response with many-body interactions using the prescriptions outlined in Refs.~\cite{maiti_etal:2016:ProbingPairingInteractionMultiple,maiti_etal:2017:ConservationLawsVertexCorrections,benek-lins_maiti:2024:ManybodyPhysicsinducedSelectionRules,sarkar_maiti:2024:ElectronicRamanResponseSuperconductor,benek-lins_maiti:2025:anisotropy_lattice}.

The relevant detail for the present work is that although we need the same steps to calculate $\chi_C$ as before, this is then followed by many recursive steps that effectively realize the many-body renormalizations. These steps involve inverting matrices of rank of the order of the mesh size. The numerical cost for such a calculation can scale uncomfortably quickly with $N$. In fact, this is one of the reasons why there are few studies exploring the Raman response from arbitrarily anisotropic superconductors~\footnote{We note that they have been addressed in the special case of a $d$-wave superconductor~\cite{devereaux_einzel:1995:ElectronicRamanScatteringSuperconductors,devereaux_einzel:1996:ErratumElectronicRamanScattering,branch_carbotte:1995:RamanElectronicContinuumSpinfluctuation,strohm_cardona:1997:ElectronicRamanScatteringCeYBa2Cu3O7}}.

In Fig.~\ref{fig:b1a1_mode}, we compare the results for the full Raman response computed with the RSum and the mRes prescriptions for different $N$ in the presence of collective modes, resulting in the presence of sharp and broad features in the converged result. The sharp one is a coherent Bardasis--Schrieffer collective mode~\cite{bardasis_schrieffer:1961:ExcitonsPlasmonsSuperconductors} of the system which provides crucial insight into the electronic properties of the material -- in this case, on the strength of sub-leading competing pairing symmetry channels~\cite{bardasis_schrieffer:1961:ExcitonsPlasmonsSuperconductors,scalapino_devereaux:2009:CollectiveWaveExcitonModes,maiti_etal:2016:ProbingPairingInteractionMultiple,benek-lins_maiti:2024:ManybodyPhysicsinducedSelectionRules} -- and is induced by many-body interactions. As is evident from the panels of the figure, for a low $N$, the distinction between the collective-mode feature and the spurious ones is not clear.
Importantly, a clear distinction is achieved faster when using the mRes prescription~\footnote{We also point the reader to a recent work, Ref.~\cite{Sondersted2024}, that discusses another method to improve computation of response functions while accounting for many-body effects.}

\subsection{Density of states}
\begin{figure*}
  \centering
  \includegraphics[width=1.0\linewidth]{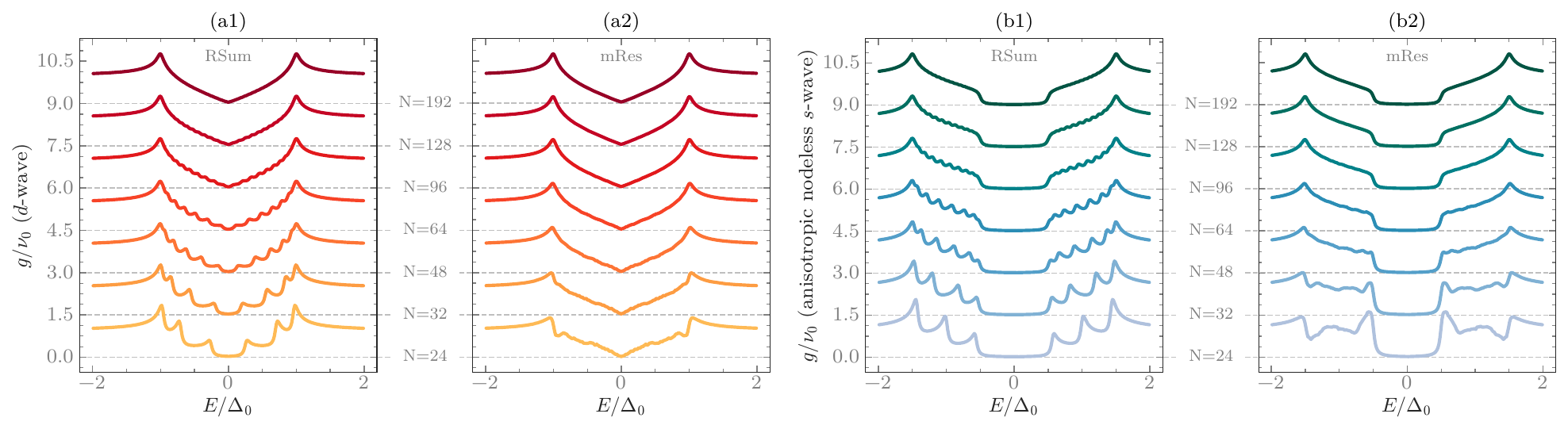}
  \caption{\textbf{Density of states.}
    Side-by-side comparison of the density of states of (a) \(d\)-wave and (b) nodeless anisotropic \(s\)-wave superconductors computed using the RSum -- (a1) and (b1) -- and the mRes -- (a2) and (b2) -- methods for different numbers of subdivisions $N$.
    In the \(d\)-wave case, the density of states for low values of \(N\) using the RSum accidentally and erroneously resembles the result for the anisotropic nodeless \(s\)-wave system. This is not the case when computing the density using the mRes prescription, which outperforms RSum in all cases.
    The densities are shifted such that the dashed horizontal lines mark the zeros.
  }\label{fig:dos}
\end{figure*}
We can provide a further example of a response function where the improvement of the results using the mRes method can also be demonstrated. Let us consider the calculation of the density of states of a superconductor with an anisotropic order parameter $\Delta_\theta$~\footnote{It really does not matter if this is for a superconductor or any other generic system. The cases of a $d$-wave and a nodeless anisotropic \(s\)-wave superconductor present opportunities to evaluate multidimensional integrals that cannot be done analytically, and hence our choice.}. The density of states per unit area of such a 2D superconductor is
\begin{align}\label{eq:DOS}
  &g(E) \nn\\
  &=\nu_0\int_0^{2\pi}\dd{\theta} \int_{0}^{\infty} \dd{\ve} \delta\left(E-\sqrt{\ve^2+\Delta_\theta^2}\right)\nn\\
  &=-\IM\Bigg(%
         \lim_{\eta\rightarrow0}\frac{\nu_0}\pi\int_0^{2\pi}\dd{\theta}\underbrace{\int_{0}^{\infty} \dd{\ve} \frac1{E-\sqrt{\ve^2+\Delta_\theta^2}+\ii\eta}}_{I(E,\theta)}
         \Bigg).
\end{align}
where $\nu_0$ is a reference density of states that is not relevant for the present discussion. The form of the singularity of this integrand is $s(E,\theta)=1/\sqrt{E-|\Delta_\theta|+\ii\eta}$ and we can straightforwardly carry out its integration by applying our mRes prescription. The results for \(d\)-wave and nodeless anisotropic \(s\)-wave superconductors are shown in \cref{fig:dos}, where the mRes method is shown to outperform the RSum. It can be seen even in this case that the RSum at low $N$ gives results that cannot distinguish between the nodal and nodeless scenarios. This is better resolved with mRes.

\section{Error analysis}\label{Sec:ErrorAnalysis}

In this section, we quantify the error in our approximation. While the improvement in the cases discussed here is visibly evident, it is important to understand the bounds within which this prescription operates. Observe that the main change of our method from the Riemann-sum prescription is that the uniform weight $\delta$ of each interval $m$ is modified as
\begin{equation}\label{eq:modify}
\delta \rightarrow \frac{S_m(\omega)}{s(\omega,\bar x_m)} \approx\begin{cases}
    \delta~&\text{if } m\neq m^*,\\
    \dfrac{S(\omega,m^*)}{s(\omega,m^*)}&\text{otherwise}.
\end{cases}
\end{equation}

As stated earlier, the prescription implements the Riemann sum in all non-singular sub-intervals and replaces the Riemann weight in the singular sub-interval with the modified-residue weight. This weight removes the singularity of $I(\omega,x)$ and leads to a finite result. A similar substitution would apply for incorporation with other integration methods. Thus, what would have conventionally required further subdivision of the sub-interval to arrive at the right result, this prescription achieves without the extra subdivisions. This is the source of the improvement.

\begin{figure*}
  \centering
  \includegraphics[width=1\linewidth]{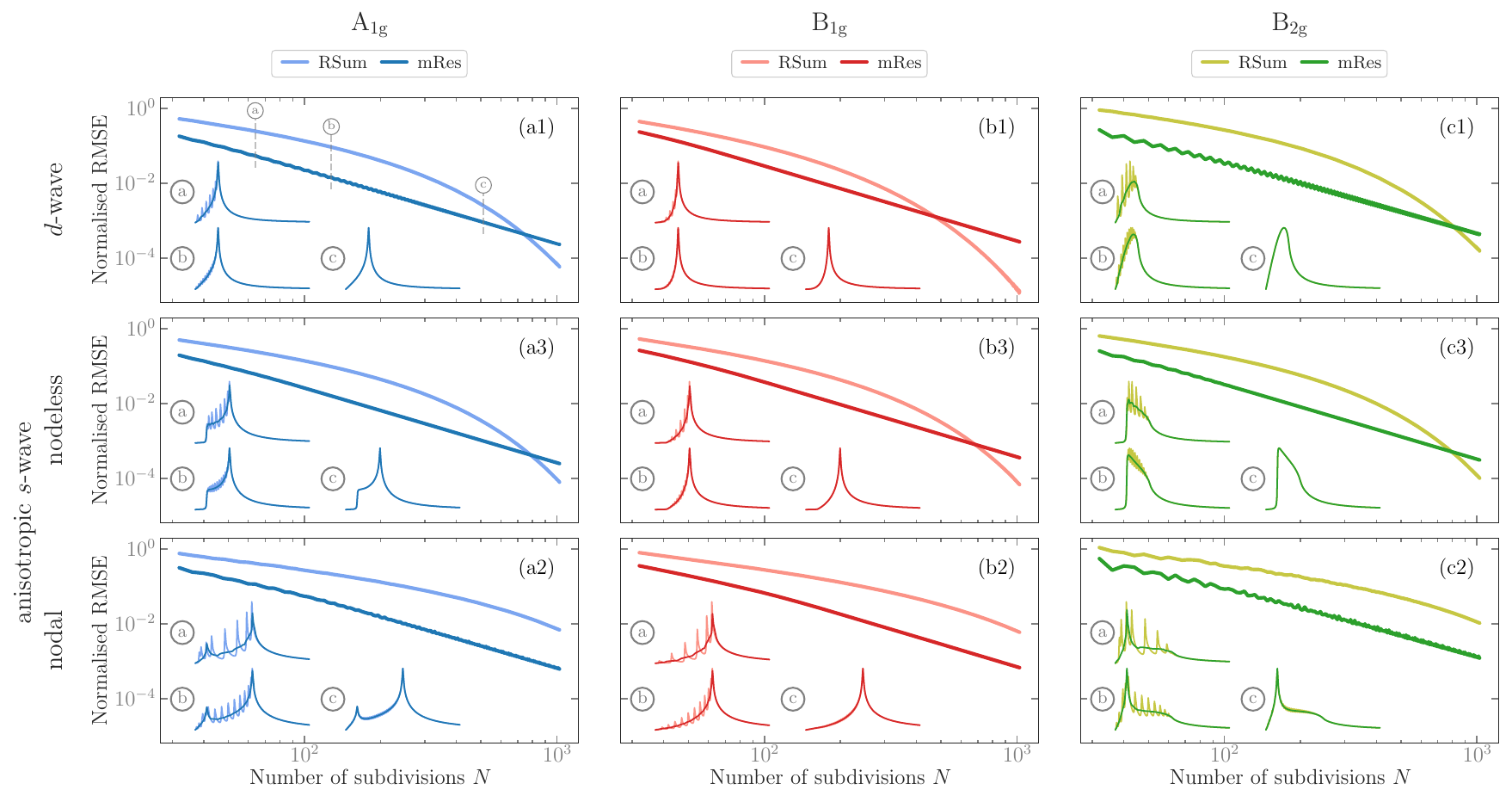}
  \caption{\textbf{Error analysis.} Normalized root mean square error (RMSE) for all the comparisons between the RSum and mRes methods presented before. The cuts $a$, $b$ and $c$ are at the same value of $N$'s in all the plots, namely $64$, $128$ and $512$. The corresponding responses are shown as insets for reference. We see that in every case the mRes approach yields a smaller RMSE for lower $N$. The non-universal critical $N$ beyond which the RSum performs better is found only for a relatively large \(N\).}\label{fig:error_analysis}

  \bigskip

  \includegraphics[width=0.667\linewidth]{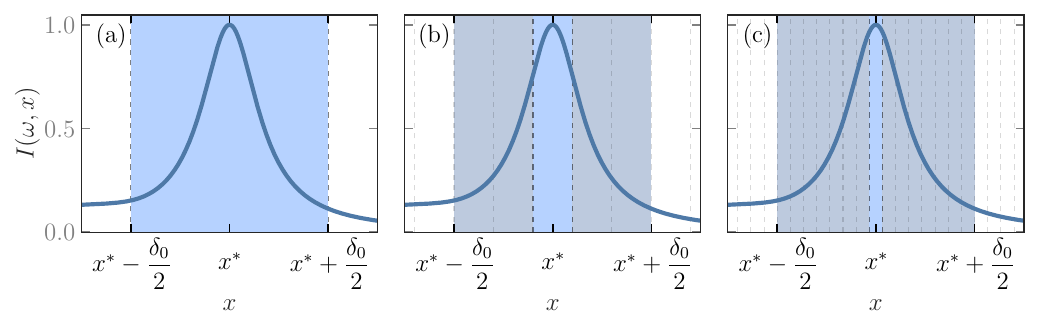}
  \caption{\textbf{A lower bound on the sub-interval size.}
    The coloured region represents the natural spread $\delta_0$ of a singularity at \(x=x^{*}\).
    This scale is non-universal and depends on the nature of the singularity.
    The blue region represents the spread of the singular sub-interval which depends on the choice of $\delta$, which is reduced across panels. When $\delta\gtrsim\delta_0$, mRes correctly integrates the singularity.
    For $\delta<\delta_0$, we introduce the grey regions that we wish to avoid. In this region the singularity is medium-varying and is not accounted for in the mRes prescription. At the same time, the narrowing of the singular sub-interval (blue) also indicates that we do not adequately pick up the correct contribution from the singularity, leading to an underperformance of the mRes prescription. The avoidance of the gray regions and the narrowing of blue regions set a lower limit to the interval size $\delta$ which are delimited here by the dashed vertical lines.}
  \label{fig:interval_interplay}
\end{figure*}
In fact, in \cref{fig:error_analysis}, we show the normalized root mean square error (RMSE) of the two methods as function of $N$ for all the cases presented in \cref{fig:ils_example,fig:ils_results,fig:ils_comparison}. Observe that for $N$ ranging from small to sufficiently large ($\sim 10^3$), we have the residue prescription outperform the Riemann sum by an order of magnitude in many cases. However, observe that there is a crossover point (that is not reached in all the panels) beyond which the Riemann sum outperforms the residue prescription. Although this happens at $N\sim 10^3$, which is already a relatively large number of sub-intervals that will seldom be necessary in practice and the error still decreases, it is incumbent on the presenters of the prescription to address this tendency. This is discussed next.

\subsection{Constraint on the sub-interval size}
We first point out that there can be two types of non-singular sub-intervals in our prescription. One where $I(\omega,x)$ is slowly varying (which is what was discussed above) and another where $I(\omega,x)$ is not singular, but medium-to-fastly varying. Ideally, if we chose an appropriate size of the sub-interval, we would not encounter these non-singular medium-fast-varying sub-intervals. But as we reduce $\delta$, we are bound to increase the number of these sub-intervals. See \cref{fig:interval_interplay} where this idea is illustrated. In these sub-intervals (shown in gray), our approximation of slow-varying $I(\omega,x)$ breaks down leading to incorrect weights for the intervals. This is the source of the prescription eventually underperforming compared to the Riemann sum for larger \(N\) -- that is, as \(\delta \to 0\).

Unfortunately, there is not a universal scheme to identify a `critical' length of the sub-interval $\delta_0$ as different singularities have different ranges over which they need to be integrated over. Fortunately, however, this is not usually a concern as it is encountered only at large $N$'s, where the use of conventional integration routines would be sufficient to obtain good results. We remind the reader that this is a prescription designed to help improve the results when the mesh is \emph{coarse}.

\subsection{The order of integration and taking imaginary part}

\begin{figure}
  \centering
\includegraphics[width=\linewidth]{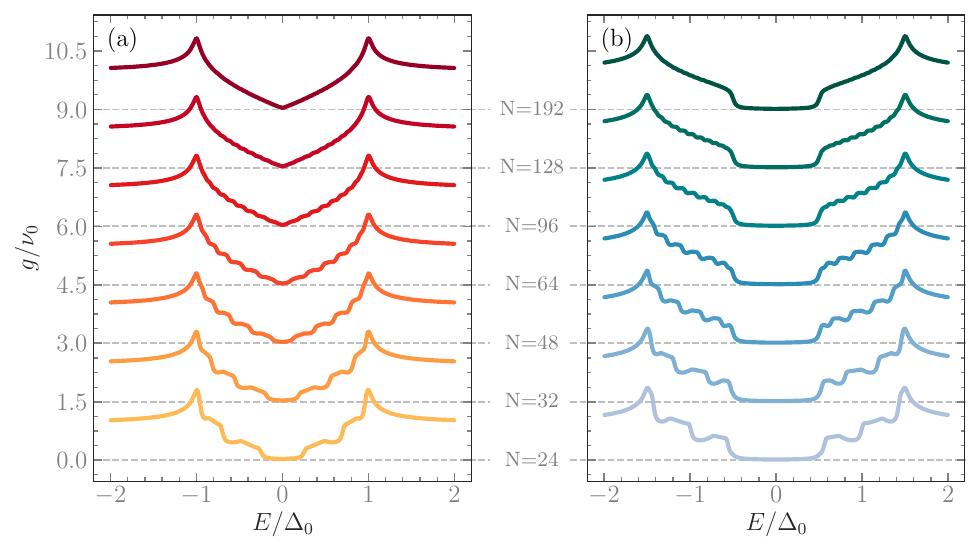}
\caption{\textbf{Order of taking the imaginary part.} Density of states of (a) \(d\)-wave and (b) nodeless anisotropic \(s\)-wave superconductors computed using the mRes prescription where the imaginary part in \cref{eq:DOS} was taken before applying the prescription.
The result should be insensitive to the order of taking the imaginary part; which they are for larger $N$, vide \customcrefs{fig:dos}{(a2)} and \hyperref[fig:dos]{(b2)}. Since \cref{eq:DOS} is $\ln$-divergent, the results for low values of \(N\) differ whilst still being an improvement over the RSum method, vide \customcrefs{fig:dos}{(a1)} and \hyperref[fig:dos]{(b1)}.
}\label{fig:dos_overall_im}
\end{figure}
We have seen in the examples above that it is the imaginary part of the integral that maps to the physical response. In principle, there is a question whether the order of taking the imaginary part and the integration are interchangeable. The short answer is yes. This is guaranteed by the fact that the regulator $\eta$ keeps everything finite, in which case the order of operations commute. Although the exact result (limit of $N\rightarrow\infty$) is the same when computed in either ordering, it requires some care when making an estimate with a discrete sum. And this is of concern to us because our objective is to achieve acceptable results in coarser grids.

To this effect, if the integral that led to $I(\omega,\theta)$ is convergent at the upper limit of the integration, then once again we would not have to worry about the order of actions. However, if the integral is divergent at the upper limit in the real part (the imaginary part still has to be convergent as it is the physical result), at low $N$'s this divergence may not effectively cancel itself out and can lead to relatively worse performance if we retain both real and imaginary parts in the prescription, taking the imaginary part only of the final result.

This can be improved by explicitly taking the imaginary part of $I(\omega,\theta)$ right from the beginning. In our example for Raman scattering $I\sim \int^\Lambda \dd{\varepsilon} /\varepsilon^2\sim1/\Lambda\rightarrow 0$ for $\Lambda\rightarrow\infty$ and, hence, the order of taking imaginary parts do not matter. However, for the example of the density of states, we have $I\sim \int^\Lambda \dd{\varepsilon} /\varepsilon\sim\ln(\Lambda)\rightarrow \infty$ for $\Lambda\rightarrow\infty$.
We exemplify this in \cref{fig:dos_overall_im}, where we show the results of computing the density of states by first taking the imaginary part of the response function and then applying the mRes method. Whilst for low \(N\)'s, the order of taking the imaginary part produces different results, vide \customcrefs{fig:dos}{(a2)} and \hyperref[fig:dos]{(b2)}, the result is still in the vicinity of the expected result and an improvement over RSum.

\section{Conclusion}\label{Sec:Conclusion}
We have presented a numerical prescription to calculate integrals that have a singular integrand, which can also result in singular features after integration. What sets it apart from other available prescriptions is that it handles and reproduces the singular features without the need of an adaptive mesh while outperforming conventional methods on a coarse grid.
This scenario is common to many practical calculations, particularly of dynamical correlation functions, where the number of subdivisions cannot be made arbitrarily small due to the computational cost of the problem.
In such scenarios, a Riemann sum prescription to carry out integrations becomes the only possibility for performing the integration.

Our prescription offers a way to better estimate the true result even with limited grid points. We demonstrated its validity in a number of scenarios. We calculated the dynamical correlation functions that are relevant for Raman scattering from superconductors for various structures of the order parameter and also to a density-of-states calculation. We demonstrated its effectiveness even in a realistic calculation of the Raman response including many-body effects that result in collective modes.

We also identified the two primary limitations of this method. First, one needs to know the nature of singularity of the integrand. The second, is that there is a lower limit to the size of the sub-interval beyond which it stops outperforming the conventional prescriptions. The former limitation is not really limiting given recent advances in theory~\cite{benek-lins_etal:2024:UniversalNonanalyticFeaturesResponse} that allow us to infer the local asymptotic forms of singularities. The latter is also shown to only be a limitation for impractically small sub-interval sizes. These considerations allow for a rather wide applicability of this technique. We also resolved a potential issue with divergent integrals where the order of taking imaginary part and doing the integration could matter.

Looking ahead, on the materials-science front, we believe that this prescription can improve analyses of anisotropic superconductors using not only Raman spectroscopy, but also infrared spectroscopy and THz pump-probe spectroscopy, where similar correlation functions are needed to model the response. One can also foresee the incorporation of this scheme into techniques based on fRG and on generalized RPA to calculate response functions after computing the ground states (which is what they are usually set-up to do). On the numerical side, one can investigate a possible incorporation into the Monte Carlo techniques of integration and possibly extend similar singularity-based integration to higher dimensional integrals.

\paragraph*{Acknowledgments:}
This work was funded (S.M.) by the Natural Sciences and Engineering Research Council of Canada (NSERC) Grant No. RGPIN-2019-05486 and partially (I.B.-L.) via the NSERC's QSciTech CREATE programme. Also, this research was enabled in part by the computing resources provided by Calcul Québec and the Digital Research Alliance of Canada. J.D.\ was in Concordia University when the initial results of the work were obtained.

\appendix
\section{Integration in the singular sub-interval}
\label{sec:appendix}

For completeness, we note the specific form of the functional \(S\) that enters the modified-residue term used to compute the bare Raman response. Specifically, from the local asymptotic form of $s(\omega,\theta)=1/\sqrt{\omega/2|\Delta_\theta|-1}$ identified in \cref{eq:I1} and for \(\eta \to 0\), we need
\begin{equation}\label{eq:app:S}
  S_{m}(\omega) = \int_{\theta_{m-1}}^{\theta_{m}} \frac{1}{\sqrt{\dfrac{\omega}{2|\Delta_{\theta}|} - 1}} \dd{\theta}.
\end{equation}

As argued in the main text, this integration is only really relevant in the singular sub-interval where $\omega\sim 2|\Delta_\theta|$. It is then desirable to remove the modulus sign by choosing $s(\omega,\theta)=1/\sqrt{[\omega/(2\Delta_\theta)]^2-1}$ and
\begin{equation}\label{eq:app:S2}
  S_{m}(\omega) = \int_{\theta_{m-1}}^{\theta_{m}}  \frac{1}{\sqrt{\left(\dfrac{\omega}{2\Delta_{\theta}}\right)^2 - 1}} \dd{\theta}.
\end{equation}

Since the interval $[\theta_{m-1},\theta_m)$ is small, one can perform the expansion $\Delta_\theta=\Delta_{\bar\theta_{m}}+v_{m}(\theta-\bar\theta_{m})$, where $v_{m}\equiv \partial_{\theta}\Delta_{\bar\theta_m}$. The above two forms of $s(\omega,\theta)$ lead to (in the end-point prescription):
\begin{widetext}
\begin{align}
    S_m(\omega)=\begin{cases}
        \dfrac{2\Delta_{\bar\theta_{m}}}{v_{m}}\left(\sqrt{A_{m}-v_{m}[\theta_{m-1}-\bar\theta_m]/\Delta_{\bar\theta_{m}}}-\sqrt{A_{m}-v_{m}[\theta_{m}-\bar\theta_m]/\Delta_{\bar\theta_{m}}}\right) &\text{for } s(\omega,\bar\theta_m)= \dfrac{1}{\sqrt{A_{m}}},\\
        \dfrac{\Delta_{\bar\theta_{m}}}{v_{m}}\left(\sqrt{A_{2,m}-2v_{m}[\theta_{m-1}-\bar\theta_m]/\Delta_{\bar\theta_{m}}}-\sqrt{A_{2,m}-2v_{m}[\theta_{m}-\bar\theta_m]/\Delta_{\bar\theta_{m}}}\right) &\text{for } s(\omega,\bar\theta_m)=\dfrac{1}{\sqrt{A_{2,m}}},
    \end{cases}
\end{align}
\end{widetext}
where $A_{m}=\omega/(2|\Delta_{\bar\theta_m}|)-1$ and $A_{2,m}=[\omega/(2\Delta_{\bar\theta_{m}})]^2-1$. This formally poses a problem near the stationary points where $v_m=0$, but the sample points can be chosen to be misaligned with the stationary points, which takes care of the numerical ambiguity.
However, observe that if $\omega\neq 2|\Delta_{\theta_m}|$ then the smallness of $\theta_m-\theta_{m-1}$ and/or $v_m$ would render both the above forms to $S_m(\omega)/s(\omega,\bar\theta_m)=[\theta_m-\theta_{m-1}]$. This demonstrates that irrespective of the choice of representation of $s(\omega,\theta)$, the intervals are always appropriately weighted.

%

\end{document}